\documentclass{article}
\usepackage{spconf,amsmath,graphicx}
\usepackage{multirow}
\usepackage{makecell}
\usepackage{bbding}
\usepackage{amssymb}
\usepackage{xcolor}
\usepackage{subcaption}
\usepackage{bibspacing}
\usepackage{hyperref}
\usepackage{marginnote}
\usepackage{bibspacing}
\usepackage{IEEEtrantools}

\def\eg{\emph{e.g. }}

\title{Adversarial Data Augmentation Using VAE-GAN for \\
Disordered Speech Recognition }
\ninept
\name{Zengrui Jin$^{1}$${\rm \!}$,${\rm\!}$ Xurong Xie$^{2}$${\rm \!}$,${\rm\!}$ Mengzhe Geng$^{1}$${\rm \!}$,${\rm\!}$ Tianzi Wang$^{1}$${\rm \!}$,${\rm\!}$ Shujie Hu$^{1}$${\rm \!}$,${\rm\!}$ Jiajun Deng$^{1}$${\rm \!}$,${\rm\!}$ Guinan Li$^{1}$${\rm \!}$,${\rm\!}$ Xunying Liu$^{1}$}
\address{
\textit{\{zrjin, mzgeng, twang, sjhu, jjdeng, gnli, xyliu\}@se.cuhk.edu.hk, xurong@iscas.ac.cn}
\\$^{1}$ The Chinese University of Hong Kong, Hong Kong SAR, China \\
	$^{2}$ Institute of Software, Chinese Academy of Sciences, China
}

\begin{document}
\bstctlcite{IEEEexample:BSTcontrol}
\setlength{\bibitemsep}{.2\baselineskip plus .05\baselineskip minus .05\baselineskip}

\maketitle
\begin{abstract}

 Automatic recognition of disordered speech remains a highly challenging task to date. 
 The underlying neuro-motor conditions, often compounded with co-occurring physical disabilities, lead to the difficulty in collecting large quantities of impaired speech required for ASR system development. 
This paper presents novel variational auto-encoder generative adversarial network (VAE-GAN) based personalized disordered speech augmentation approaches that simultaneously learn to encode, generate and discriminate synthesized impaired speech. 
Separate latent features are derived to learn dysarthric speech characteristics and phoneme context representations. 
Self-supervised pre-trained Wav2vec 2.0 embedding features are also incorporated. 
Experiments conducted on the UASpeech corpus suggest the proposed adversarial data augmentation approach consistently outperformed the baseline speed perturbation and non-VAE GAN augmentation methods with trained hybrid TDNN and End-to-end Conformer systems.
After LHUC speaker adaptation, the best system using VAE-GAN based augmentation produced an overall WER of 27.78\% on the UASpeech test set of 16 dysarthric speakers, and the lowest published WER of 57.31\% on the subset of speakers with ``Very Low'' intelligibility.
\end{abstract}

\section{Introduction}

Despite the rapid progress in automatic speech recognition (ASR) system development targeting normal speech in recent decades, 
accurate recognition of disordered speech remains a highly challenging task to date \cite{christensen2012comparative, christensen2013combining, sehgal2015model, yu2018development, liu2020exploiting, hu2022exploiting, geng2021investigation, xie2022variational}.
Dysarthria is a common form of speech disorder caused by motor control conditions including cerebral palsy, amyotrophic lateral sclerosis, stroke and traumatic brain injuries \cite{lanier2010speech}. 
Disordered speech brings challenges on all fronts to current deep learning based ASR systems predominantly targeting normal speech recorded from healthy speakers.
First, a large mismatch between such data and normal speech is often observed. 
Second, the physical disabilities and mobility limitations often observed among impaired speakers lead to the difficulty in collecting large quantities of their speech required for ASR system development.

To this end, data augmentation techniques play a vital role in addressing the above data scarcity issue. 
Data augmentation techniques have been widely studied for speech recognition targeting normal speech.
By expanding the limited training data using, for example, tempo, vocal tract length or speed perturbation \cite{verhelst1993overlap, kanda2013elastic, jaitly2013vocal, ko2015audio}, stochastic feature mapping \cite{cui2015data}, simulation of noisy and reverberated speech to improve environmental robustness \cite{ko2017study} and back translation in end-to-end systems \cite{hayashi2018back}, the coverage of the augmented training data and the resulting speech recognition systems' generalization can be improved.

\begin{figure*}[t]
	\centering
	\includegraphics[width=1.\linewidth]{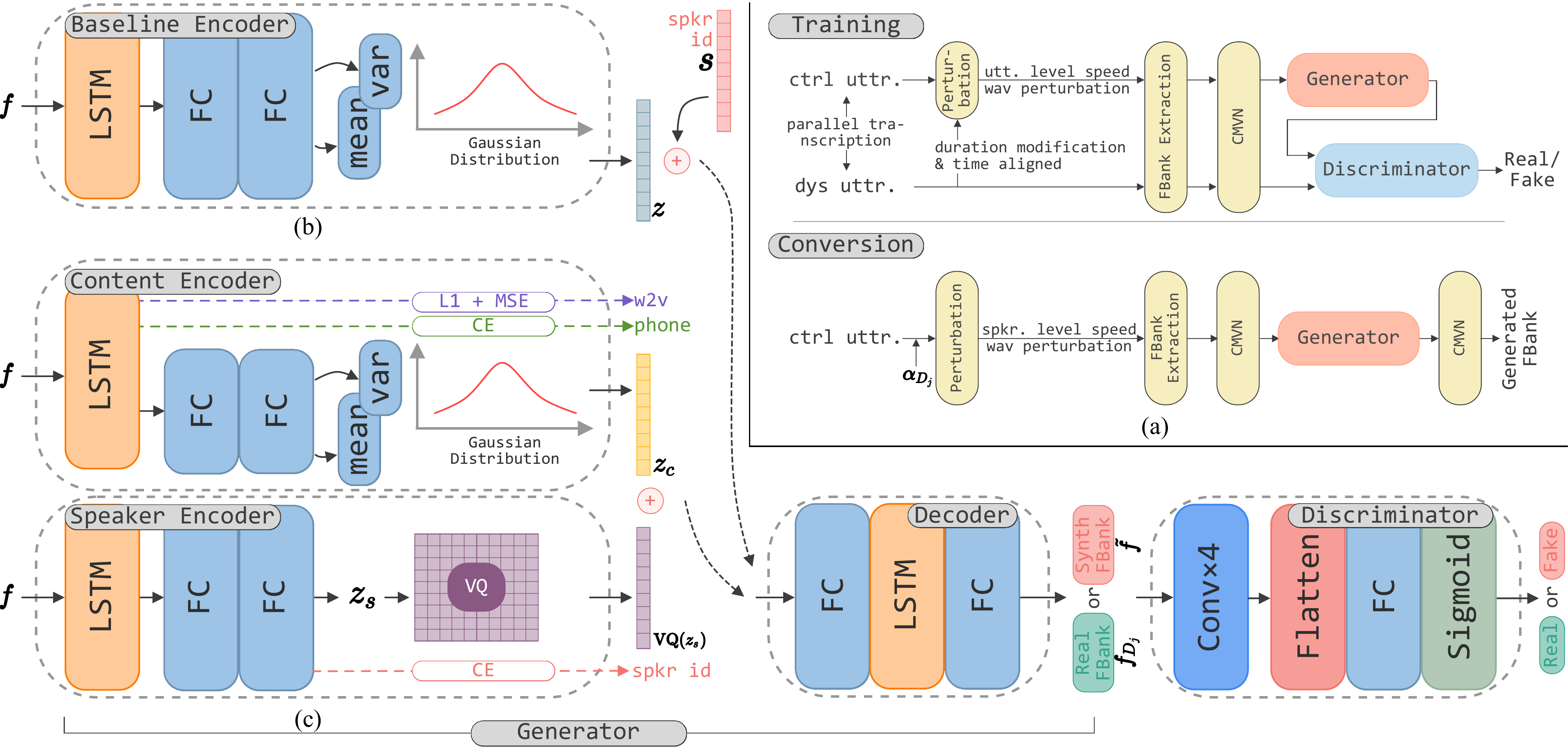}
	\caption{Example of (a) baseline non-VAE GAN based data augmentation (DA); (b) VAE-GAN based DA; (c) structured VAE-GAN based DA with its semi-supervised encoder producing separate latent content and VQ quantized speaker representations; it is trained on a combination of unsupervised features reconstruction cost in initialization, and followed by additional supervised training based on phone labels CE (green), Wav2vec 2.0 (w2v) pre-trained features L1+MSE (purple) and speaker ID CE (red) costs. $\boldsymbol{\oplus}$ denotes vector concatenation.}
	\label{fig:vae_gan_model}
\end{figure*}

With the successful application of generative adversarial networks (GANs) \cite{goodfellow2014generative} to a wide range of speech processing tasks including, but not limited to, 
speech synthesis \cite{beck2022wavebender},
voice conversion \cite{wang2021enriching}, 
speech enhancement \cite{su2021hifi}, 
speech emotion recognition \cite{eskimez2020gan}
and speaker verification \cite{du2021synaug}, 
GAN-based data augmentation methods for ASR-related applications targeting normal speech have also been studied for 
robust speech recognition \cite{chen2022noise,mirheidari2020data}, 
and whisper speech recognition \cite{gudepu2020whisper}.

In contrast, only limited research has been conducted on GAN-based data augmentation methods for disordered speech recognition tasks. 
In addition to only modifying the overall spectral shape and speaking rate as considered in widely used tempo or speech perturbation \cite{verhelst1993overlap,ko2015audio} based data augmentation techniques, GAN-based disordered speech augmentation \cite{jin21_interspeech,harvill2021synthesis} is proposed to capture more detailed spectra-temporal differences between normal and impaired speech. 
However, when modeling the rich taxonomy of heterogeneity of disordered speech, these existing GAN-based augmentation methods suffer from a lack of structured representations encoding speaker induced variability, and those associated with speaker independent temporal speech contexts. 
This drawback limits their controllability over the augmented data and generalization to diverse speakers with varying speech pathology severity levels.

To this end, a novel solution considered in this paper is to draw strengths from both variational auto-encoders (VAE) \cite{kingma2013auto,van2017neural,xie2022variational} and GANs.
This leads to their combined form, VAE-GAN \cite{larsen2016autoencoding}, which is designed to simultaneously learn to encode, generate and discriminate synthesized speech data samples. 
Inside the proposed VAE-GAN model, separate latent neural features are derived using a semi-supervised VAE component to learn dysarthric speech characteristics and phoneme context embedding representations respectively. 
Self-supervised pre-trained Wav2vec 2.0 \cite{baevski2020wav2vec} embedding features are also incorporated. 
This resulting VAE-GAN based augmentation performs personalized dysarthric speech augmentation for each impaired speaker using data collected from healthy control speakers. 

Experiments conducted on the largest publicly available UASpeech dysarthric speech corpus \cite{kim2008dysarthric} suggest the proposed VAE-GAN based adversarial data augmentation approach consistently outperforms the baseline speed perturbation and non-VAE GAN augmentation methods on state-of-the-art hybrid TDNN and Conformer end-to-end systems. 
After LHUC speaker adaptation, the best system using VAE-GAN based augmentation produced an overall WER of 27.78\% on the UASpeech test set of 16 dysarthric speakers, and the lowest state-of-the-art WER of 57.31\% on the subset of speakers with ``Very Low'' intelligibility.

The major contributions of this paper are listed below. 
First, to the best of our knowledge, this paper presents the first use of VAE-GAN based data augmentation approaches for disordered speech recognition.
In contrast, VAEs have been primarily used separately and independent of GANs across a wide range of speech processing tasks, including, but not limited to, 
speech enhancement \cite{ho22_interspeech, pariente19_interspeech}, 
speaker verification \cite{zhang19n_interspeech, vinals18_interspeech},
speech synthesis and voice conversion \cite{guo22d_interspeech, lu21d_interspeech, cao20b_interspeech}, 
and speech emotion recognition \cite{liu2020temporal, yang2019attribute}.
Their combination with GANs in the context of data augmentation for disordered speech recognition has not been studied.
Second, the final system constructed using the best VAE-GAN based augmentation approach in this paper produced the lowest published WERs of 57.31\% and 28.53\% on the ``Very Low'' and ``Low'' intelligibility speaker groups respectively on the UASpeech task published so far.

The rest of this paper is organized as follows. Disordered speech data augmentation methods based on conventional non-VAE GANs are presented in Section 2. Section 3 proposes VAE-GAN based disordered speech data augmentation. Section 4 presents the experiments and results on the UASpeech dysarthric speech corpus. The last section concludes and discusses possible future works. 

\section{Adversarial Data Augmentation for Disordered Speech Recognition}
\label{sec:baseline}

This section reviews the existing GAN-based data augmentation (DA) for disordered speech recognition \cite{jin21_interspeech,harvill2021synthesis}. 
In contrast to only modifying the speaking rate or overall shape of spectral contour of normal or limited impaired speech training data during augmentation, fine-grained spectro-temporal characteristics such as articulatory imprecision, decreased clarity, breathy and hoarse voice as well as increased dysfluencies are further injected into the data generation process during GAN training using the temporal or speed perturbed normal-dysarthric parallel data \cite{jin21_interspeech}.

\begin{table*}[!t]
\centering
\caption{Performance of TDNN systems on the 16 UASpeech dysarthric speakers using different data augmentation methods. 	``CTRL'' in the ``Data Augmentation'' column stands for dysarthric speaker dependent transformation of control speech during data augmentation using from left to right: speed perturbation in ``S''; ``SG'' denotes non-VAE GAN models \cite{jin21_interspeech}; and ``VG'' denotes structured VAE-GAN models of Sec. \ref{sec:method}. ``VL/L/M/H'' refers to intelligibility subgroups (Very Low/Low/Medium/High). ``DYS'' column denotes speaker independent speed perturbation of disordered speech. $\dag$ and $\star$ denote statistically significant improvements ($\alpha = 0.05$) are obtained over the comparable baseline systems with speed perturbation (Sys. 2, 5, 8) or speed-GAN (Sys. 3, 6, 9) respectively. }
\label{tab:overall}
\resizebox{1.\linewidth}{!}{
\begin{tabular}{c|c|c|c|c|c|c|c|c|c|c||c|c|c|c|c}
	\hline
	\hline
\multirow{3}{*}{Sys.} 
	& \multicolumn{4}{c|}{Data Augmentation} 
	& \multirow{3}{*}{\# Hrs} 
	& \multicolumn{5}{c|}{WER \% (Unadapted)} 
	& \multicolumn{5}{c}{WER \% (LHUC-SAT Adapted)} \\ 
	\cline{2-5} 
	\cline{7-16}
  & \multicolumn{3}{c|}{CTRL} 
  	& DYS 
  	&  
  	& \multirow{2}{*}{VL} 
  	& \multirow{2}{*}{L} 
  	& \multirow{2}{*}{M} 
  	& \multirow{2}{*}{H} 
  	& \multirow{2}{*}{Avg.} 
  	& \multirow{2}{*}{VL} 
  	& \multirow{2}{*}{L} 
  	& \multirow{2}{*}{M} 
  	& \multirow{2}{*}{H} 
  	& \multirow{2}{*}{Avg.} \\ 
	\cline{2-5}
  & S & SG & VG & S &  &  &  &  &  &  &  &  &  &  & \\ 
    \hline
    \hline
1 & \multicolumn{3}{c|}{-} & - & 30.6 & 70.78 & 42.82 & 36.47 & 25.86 & 41.81 & 65.78 & 38.47 & 33.27 & 23.74 & 38.29 \\
    \hline
    \hline    
2 & 1x & & & \multirow{3}{*}{-} & \multirow{3}{*}{50.2} & 64.22 & 32.13 & 23.06 & \textbf{13.13} & 30.77 & 59.39 & 29.79 & 21.94 & 13.94 & 29.16 \\
3 &  & 1x &  &  &  & \textbf{62.89}$^\dag$ & \textbf{32.06} & \textbf{22.67} & 13.88 & 30.62 & \textbf{57.20}$^\dag$ & 28.83 & 20.98 & 14.15 & 28.51$^\dag$ \\
4 &  &  &  1x &  &  & 63.37$^\dag$  & 32.14 & 23.00 & 13.18$^\star$ & \textbf{30.61} & 58.84 & \textbf{27.86}$^{\dag\star}$  & \textbf{19.73}$^{\dag\star}$  & \textbf{13.13}$^{\dag\star}$  & \textbf{27.89}$^{\dag\star}$  \\
    \hline
5 & 1x & & & \multirow{3}{*}{2x} & \multirow{3}{*}{87.5} & 62.76 & 31.71 & 24.16 & 13.62 & 30.72 & 61.09 & 29.06 & 21.14 & 12.66 & 28.76 \\
6 &  & 1x &  &  &  & \textbf{60.86}$^\dag$ & \textbf{30.81}$^\dag$ & \textbf{22.86}$^\dag$ & \textbf{13.14}$^\dag$ & \textbf{29.65}$^\dag$ & \textbf{57.97}$^\dag$ & 29.38 & 20.84 & 12.86 & 28.15$^\dag$ \\
7 &  &  & 1x &  &  & 62.10$^\dag$ & 31.21 & 23.43 & 14.02 & 30.46$^\dag$ & 60.73 & \textbf{28.77} & \textbf{18.63}$^{\dag\star}$ & \textbf{11.84}$^{\dag\star}$ & \textbf{27.88}$^\dag$ \\
    \hline
8 & 2x & & & \multirow{3}{*}{2x} & \multirow{3}{*}{130.1} & 62.55 & 31.97 & 23.12 & \textbf{13.13} &  30.56 & 60.23 & 29.02 & 20.12 & \textbf{12.52} & 28.36 \\
9 &  & 2x &  &  &  & 60.30$^\dag$  & 31.49 & 23.16 & 13.62 & 29.92$^\dag$  & 57.84$^\dag$ & 29.66 & 20.45 & 12.89 & 28.09$^\dag$ \\
10 &  &  & 2x &  &  & \textbf{59.52}$^\dag$  & \textbf{31.34} & \textbf{22.84} & 13.62 & \textbf{29.71}$^\dag$  & \textbf{57.31}$^\dag$ & \textbf{28.53}$^\star$ & \textbf{20.10}$^{\dag\star}$ & 13.04 & \textbf{27.78}$^\dag$ \\ 
    \hline
    \hline
\end{tabular}
}
\end{table*}

An example of such GAN base DA approach is shown in Fig. \ref{fig:vae_gan_model}(a) (top right). 
During the training stage, for each dysarthric speaker, the duration of normal, control speech utterances is adjusted via waveform level speed or temporal perturbation \cite{ko2015audio, verhelst1993overlap} and time aligned with disordered utterances containing the same transcription. 
Mel-scale filter-bank (FBank) features extracted from these parallel utterance pairs are used to train speaker dependent (SD) GAN models. 
During data augmentation, the resulting speaker specific GAN models are applied to the FBank features extracted from speaker level perturbed normal speech utterances to generate the final augmented data for each target impaired speaker. 
The resulting data are used together with the original training set during ASR system training. 
Without incorporating any VAE component in the generator, such GAN model serves as the baseline adversarial data augmentation approach of this paper. 
The architecture of the baseline non-VAE GAN follows the hyper-parameter configuration of the Encoder, Decoder and Discriminator network in the top left sub-figure (b) and the bottom right corners of Fig. \ref{fig:vae_gan_model}, but without the VAE latent feature Gaussian distribution. 
On top of such baseline GAN architecture, a series of VAE-GAN models are introduced in Sec. \ref{sec:method} below by replacing the non-VAE GAN generator with various forms of VAEs, while the overall data preparation, model training and data generation procedures remain the same.

Letting $\boldsymbol{f}=\{f_{t=1:T}\}$ denote an acoustic feature sequence, the general GAN training objective function both maximizes the binary classification accuracy on target disordered speech and minimizes that obtained on the GAN perturbed normal speech. It is expected that upon convergence, the latter is modified to be sufficiently close to the target impaired speech. This is given by

\begin{align}
	\mathop{\rm min}\limits_{G_j}^{}\mathop{\rm max}\limits_{D_j}^{} &\ V(D_j, G_j) \\
	&\ = \mathbb{E}_{\boldsymbol{f}_{D}\sim p_{D_j}(\boldsymbol{f})}[\log{(D_j(\boldsymbol{f}_{D_j}))}] \\
	&\ + \mathbb{E}_{\boldsymbol{f}_{C} \sim p_{C}(\boldsymbol{f})}[\log{(1-D_j(G_j(\boldsymbol{f}_{C})))}]
\label{eq:ori_gan}
\end{align}

where $j$ represents the index for each target dysarthric speaker, $G_j$ and $D_j$ are the Generator and Discriminator associated with dysarthric speaker $j$, $\boldsymbol{f}_C$ and $\boldsymbol{f}_{D_j}$ stand for the FBank features of paired control and dysarthric utterances.

\section{VAE-GAN Based Data Augmentation}
\label{sec:method}

This section presents VAE-GAN based data augmentation approaches. The generator module is constructed using either a standard VAE, or structured VAE producing separate latent content and speaker representations with additional Wav2vec 2.0 SSL pre-trained speech contextual features incorporated.

\subsection{Standard VAE-GAN}
\label{sec:standard_vae_gan}
The overall architecture of a standard VAE-GAN consists of three components: an Encoder, a Decoder and a Discriminator, respectively shown in the top left sub-figure (b) and the bottom right corners of Fig. \ref{fig:vae_gan_model}. 
The Encoder and Decoder together form the GAN generator. 
During the training stage, the VAE-based generator is initialized using the original, un-augmented UASpeech training data (with silence stripping applied) in an unsupervised fashion.

On the detailed hyper-parameter configurations, the Encoder network (Fig. \ref{fig:vae_gan_model}(b)) consists of one LSTM network with input and hidden size set to $40$ and $128$ respectively, followed by two fully connected (FC) layers, both with $128$-dim input and output.
A Gaussian variational distribution is formed using mean and covariance matrices produced by two individual fully connected layers with an input size of $128$. 
$39$-dimensional encoded latent features $\boldsymbol{z}$ are sampled from this Gaussian distribution before being fed into the Decoder network for feature reconstruction. 
The Decoder (Fig. \ref{fig:vae_gan_model}, bottom right) consists of the following component layers applied in sequence: an FC layer with an output size of $128$; an LSTM layer with the input and hidden sizes set to $40$ and $128$; and a 3$^{rd}$ FC layer with its input and output size set to $128$ and $40$. 
The VAE generator network is optimized to minimize the KL divergence between the probabilistic encoder output distribution $q(\boldsymbol{z}|\boldsymbol{f}, \boldsymbol{\phi})$ and the prior distribution $p(\boldsymbol{z})$ via the following lower bound,

\begin{align}
	\log &\ p(\boldsymbol{f}|\boldsymbol{\theta}) = \log \int p(\boldsymbol{f},\boldsymbol{z} | \boldsymbol{\theta}) d \boldsymbol{z} \geq \\
	&\ \int q(\boldsymbol{z}|\boldsymbol{f}, \boldsymbol{\phi}) \log p(\boldsymbol{f} | \boldsymbol{z'},\boldsymbol{\theta})d \boldsymbol{z} - \mathbb{KL}(q||p) \overset{\rm def}{=}L_{\rm vlb}
    \label{eq:vae_sum_loss_func}
\end{align}

$\boldsymbol{\theta}$ and $\boldsymbol{\phi}$ represent parameters of decoder and encoder respectively, $\boldsymbol{z'} = \boldsymbol{z} \oplus \boldsymbol{s}$, concatenated with the one-hot speaker ID vector $\boldsymbol{s}$ associated with each control or dysarthric speaker. 

After the initialization stage, in common with the baseline GAN of Sec. \ref{sec:baseline}, the VAE-based generator (Fig. \ref{fig:vae_gan_model}(b)) is jointly trained together with the discriminator (Fig. \ref{fig:vae_gan_model}, bottom right) in an adversarial manner using GAN Loss on the speed perturbed parallel normal-dysarthric speech utterances until convergence.

\subsection{Structured VAE-GAN}
\label{sec:structured_vae_gan}

To further enhance the controllability in personalized data augmentation for each impaired speaker during ASR system training, structured VAE-based generator producing separate latent content and speaker representations is designed and shown in the ``Content Encoder'' and ``Speaker Encoder'' blocks of Fig. \ref{fig:vae_gan_model}(c). 
During the initialization stage before later connected with the Decoder and Discriminator in joint adversarial estimation, both the content and speaker encoders are trained jointly using the variational inference lower bound of Sec. \ref{sec:standard_vae_gan}. 
Additional supervised mono-phone (green dotted line, Fig. \ref{fig:vae_gan_model}(c)) and speaker ID classification (red dotted line, Fig. \ref{fig:vae_gan_model}(c)) CE error costs are used in the content and speaker encoders respectively. 
To produce the most distinguishable speaker level latent features, $29$-dimensional vector code book quantization of the speaker encoder outputs is also applied (purple, Fig. \ref{fig:vae_gan_model}(c)). 
The size of the speaker VQ code book is $29$. All the other hyper-parameters in the other encoder submodules remain the same as the standard VAE-GAN in Fig. \ref{fig:vae_gan_model}(b). 
The variational lower bound of standard VAE in Eqn. (\ref{eq:vae_sum_loss_func}) is modified as
\begin{align}
	&\ \log p(\boldsymbol{f}|\boldsymbol{\theta}) = \log \int p(\boldsymbol{f},\boldsymbol{z''} | \boldsymbol{\theta}) d \boldsymbol{z''} \geq \\
	&\ \int q_{\rm cnt}(\boldsymbol{z}_{\rm c}|\boldsymbol{f}, \boldsymbol{\phi}_{\rm cnt}) q_{\rm spkr}(\boldsymbol{z}_{\rm s}|\boldsymbol{f}, \boldsymbol{\phi}_{\rm spkr}) \log p(\boldsymbol{f} | \boldsymbol{z''},\boldsymbol{\theta})d \boldsymbol{z''} \\
	&\ \qquad \qquad \qquad \qquad - \mathbb{KL}(q_{\rm cnt}||p) - \mathbb{KL}(q_{\rm spkr}||p) \overset{\rm def}{=}L_{\rm vlb}
\end{align}

where the structured variational distributions $q_{\rm cnt}(\boldsymbol{z}_{\rm c} |\boldsymbol{f}, \phi_{\rm cnt})$ and $q_{\rm spkr}(\boldsymbol{z}_{\rm s} |\boldsymbol{f}, \phi_{\rm spkr})$ are used by the Content and Speaker Encoders respectively, and the latent feature vector $\boldsymbol{z''} = \boldsymbol{z}_{\rm c} \oplus {\rm VQ}(\boldsymbol{z}_{\rm s})$. ${\rm VQ}(\boldsymbol{z_s})$ is the quantized vector given by $\underset{{\rm VQ}(\boldsymbol{z_s})}{\rm argmin}(L_2({\boldsymbol{z_s}}, {\rm VQ}(\boldsymbol{z_s}))$.
To help generate structured latent features, extra supervision is added during the training process as following

\begin{align}
	L_{\rm cnt}& =  \alpha L_{\rm CE}^{\rm phn} \\
	L_{\rm spkr}& = \beta L_{\rm CE}^{\rm spkr} + \gamma L_{\rm VQ} 
\label{eq:structured_vae_loss_func}
\end{align}

where $\alpha$, $\beta$ and $\gamma$ are empirically set to $1$, $1$ and $0.2$ respectively. The VQ quantization error cost $L_{\rm VQ}$ is 

\begin{align}
	L_{\rm VQ} = ||\boldsymbol{z_s} - {\bf sg}[{\rm VQ}(
	\boldsymbol{z_s})]||^2_2 + ||{\bf sg}[\boldsymbol{z_s}]-{\rm VQ}(
	\boldsymbol{z_s})||^2_2
 \label{eq:vq_loss}
\end{align}

$\boldsymbol{z_s}$ and ${\bf sg}(\cdot)$ stands for the hidden layer feature of the speaker encoder and ``stop gradient'' operator (used in error forwarding only). 
The resulting $39$ and $29$ dimensional content and speaker VAE latent representations $\boldsymbol{z_c}$ and ${\rm VQ}(\boldsymbol{z_s})$ are then concatenated before being fed into the Decoder during joint adversarial training. 

During the data generation, the VQ quantized latent speaker representations ${\rm VQ}(\boldsymbol{z_s})$ learned in VAE-GAN training for each dysarthric speaker is fixed, while the source control speaker's data is feed-forwarded through the VAE Content Encoder to produce the content representations $\boldsymbol{z_c}$, before them being concatenated to produce the synthesized impaired speech for the target speaker.

\subsection{Structured VAE-GAN with Wav2vec 2.0 Features}

To the quality of the latent content feature extraction, $256$-dimensional self-supervised pre-trained speech representation sequences produced by the Wav2vec 2.0 \cite{baevski2020wav2vec} model's bottleneck layer after being fine-tuned on the UASpeech dataset are incorporated as an additional $L_1$ plus $\rm MSE$ cost based regression task during the content encoder initialization (purple dotted line, Fig. \ref{fig:vae_gan_model}(c)).  

\section{Experiments and Results}

\begin{table}[!t]
\centering
\caption{Ablation study conducted on augmented UASpeech 130.1 hours training set. ``TDNN X$\rightarrow$Y'' denotes TDNN system X produced N-best outputs in the $1^{st}$ decoding pass prior to $2^{nd}$ pass rescoring by Sys. Y using cross-system score interpolation \cite{cui22_interspeech}.}
\renewcommand\tabcolsep{2.7pt}
\label{tab:ablation_study}
\resizebox{1.\linewidth}{!}{
	\begin{tabular}{c|c|ccc|c|c|c|c|c|c}
	\hline
	\hline
		\multirow{2}{*}{Sys.} & \multirow{2}{*}{\makecell[c]{Model\\(\# Param.)}} & \multicolumn{3}{c|}{Supervision} & \multirow{2}{*}{\makecell[c]{LHUC\\SAT}} & \multicolumn{5}{c}{WER \%} \\
		\cline{3-5}
		\cline{7-11}
		 & & spkr & phone & w2v &  & VL & L & M & H & Avg. \\
		 \hline
		 1 & \multirow{10}{*}{\makecell[c]{Hybrid\\TDNN\\(6M)}} & \multicolumn{3}{c|}{-} & \multirow{5}{*}{-} & 61.50 & 32.04 & 22.84 & \textbf{13.61} & 30.30 \\
		 2 & & \checkmark & & & & 61.28 & 30.99 & \textbf{22.41} & 13.69 & 29.93 \\
		 3 & & \checkmark & \checkmark & & & 61.89 & 31.34 & 23.51 & 13.65 & 30.34 \\
		 4 & & \checkmark & & \checkmark & & 62.34 & \textbf{30.62} & 23.25 & 13.76 & 30.24 \\
		 5 & & \checkmark & \checkmark & \checkmark & & \textbf{59.52} & 31.34 & 22.84 & 13.62 & \textbf{29.71} \\
		 \cline{1-1}
		 \cline{3-11}
		 6 & & \multicolumn{3}{c|}{-} & \multirow{5}{*}{\checkmark} & 61.00 & 29.24 & 19.67 & 13.03 & 28.66 \\
		 7 & & \checkmark & & & & 59.39 & 29.28 & 19.06 & \textbf{11.84} & 27.82 \\
		 8 & & \checkmark & \checkmark & & & 60.37 & 28.64 & \textbf{18.90} & 12.26 & 27.97 \\
		 9 & & \checkmark & & \checkmark & & 60.50 & \textbf{28.29} & 19.27 & 12.29 & 27.99 \\
		 10 & & \checkmark & \checkmark & \checkmark & & \textbf{57.31} & 28.53 & 20.10 & 13.04 & \textbf{27.78} \\
		 \hline
		 \hline
		 11 & \multirow{5}{*}{\makecell[c]{Conformer\\+SpecAug\\(40M)}} & \multicolumn{3}{c|}{-} & \multirow{5}{*}{-} & 84.81 & 58.48 & 49.33 & 38.22 & 55.47 \\
		 12 & & \checkmark & & & & 76.29 & 52.02 & 47.27 & 37.16 & 51.24 \\
		 13 & & \checkmark & \checkmark & &  & 75.77 & 51.34 & 47.03 & 37.91 & 51.16 \\
		 14 & & \checkmark & & \checkmark &  & 84.59 & 57.64 & 48.54 & 36.97 & 54.64 \\
		 15 & & \checkmark & \checkmark & \checkmark &  & \textbf{75.01} & \textbf{45.35} & \textbf{43.77} & \textbf{35.01} & \textbf{47.34} \\
		 \hline
		 \hline
		 16 & TDNN 6$\rightarrow$11 & \multicolumn{3}{c|}{-} & \multirow{5}{*}{-} & 59.63 & 28.87 & 19.51 & 12.13 & 27.94 \\
		 17 & TDNN 7$\rightarrow$12 & \checkmark &  &  &  & 59.25 & 29.25 & 19.00 & \textbf{11.56} & 27.67 \\
		 18 & TDNN 8$\rightarrow$13 & \checkmark & \checkmark &  &  & 60.23 & 28.69 & \textbf{18.75} & 11.89 & 27.79 \\
		 19 & TDNN 9$\rightarrow$14 & \checkmark &  & \checkmark &  & 60.59 & 29.24 & 19.71 & 11.87 & 28.19 \\
		 20 & TDNN 10$\rightarrow$15 & \checkmark & \checkmark & \checkmark &  & \textbf{58.31} & \textbf{28.55} & 19.88 & 11.91 & \textbf{27.58} \\
		 \hline
		 \hline
	\end{tabular}
}
\end{table}

Experiments are conducted on the UASpeech database \cite{kim2008dysarthric}, the single word dysarthric speech recognition task. 
It contains 102.7 hours of speech from 16 dysarthric and 13 control speakers with 155 common and 300 uncommon words.
Speech utterances are divided into 3 blocks, each containing all the common words and one-third of the uncommon words.
Block 1 and 3 serve as the training data while block 2 of the 16 dysarthric speakers is used as the test set. 
Silence stripping using GMM-HMM systems follows our previous work \cite{liu2020exploiting}. 
Without data augmentation (DA), the final training set contains 99195 utterances, around 30.6 hours. 
The test set contains 26520 utterances, around 9 hours.

\subsection{Experimental Setup}
The proposed VAE-GAN model is implemented with PyTorch.
The hybrid system is an LF-MMI factored time delay neural network (TDNN) system \cite{peddinti2015time, povey2016purely} containing $7$ context slicing layers trained following the Kaldi chain system setup, except that i-Vector features are not incorporated.
300-hr Switchboard data pre-trained Conformer ASR systems are implemented using the ESPnet toolkit \footnote{12 encoder layers + 6 decoder layers, feed-forward layer dim = 2048, attention heads = 4, attention heads dim = 256, interpolated CTC+AED cost.} to directly model grapheme (letter) sequence outputs, before being domain fine-tuned to the UASpeech training data. 
We use HTK toolkit \cite{young2002htk} for phonetic analysis, silence stripping and feature extraction.
Speed perturbation is implemented using SoX\footnote{SoX, audio manipulation tool. Available at: \url{https://sox.sourceforge.net}}.

\subsection{Performance of VAE-GAN DA}
Tab. \ref{tab:overall} (col. 7 - 11) presents the performance of speaker independent (SI) TDNN systems with different data augmentation methods prior to applying LHUC-SAT \cite{swietojanski2016learning} speaker adaptation (last 5 col.).
Several trends can be observed from Tab. \ref{tab:overall}: 
\textbf{1)} Across different quantities of augmented training data, the proposed VAE-GAN approach consistently outperforms speed perturbation on the ``Very Low'' (VL) and ``Low'' (L) intelligibility subgroups by up to 3.03\% absolute (4.84\% relative) WER reduction (\eg Sys. 10 \textit{vs.} 8, col. 7 for ``VL''); 
\textbf{2)} Our proposed VAE-GAN approach outperforms the non-VAE GAN approach (\eg Sys. 10 \textit{vs.} 9, col. 7 \& 8) on the ``VL'' and ``L'' intelligibility subgroups by up to 0.78\% absolute (1.29\% relative) WER reduction. 

\begin{table}[!t]
	\centering
	\caption{A comparison between published systems on UASpeech and our system. ``DA'' stands for data augmentation. ``L'', ``VL'' and ``Avg.'' represent WER (\%) for low, very low intelligibility group and average WER.}
	\label{tab:comparison}
	\renewcommand\arraystretch{1.0}
	\resizebox{1.\linewidth}{!}{
	\begin{tabular}{c|c|c|c}
	    \hline
	    \hline
		Systems & VL & L & Avg. \\
		\hline
		CUHK-2018 DNN System Combination \cite{yu2018development} & - & - & 30.60 \\
		Sheffield-2019 Kaldi TDNN + DA \cite{xiong2019phonetic} & 67.83 & 27.55 & 30.01 \\
		\makecell{Sheffield-2020  CNN-TDNN speaker adaptation \cite{xiong2020source}} & 68.24 & 33.15 & 30.76 \\
		CUHK-2020 DNN + DA + LHUC-SAT \cite{geng2021investigation} & 62.44 & 27.55 & 26.37 \\
		CUHK-2021 LAS + CTC + Meta Learning + SAT \cite{wang2021improved} & 68.70 & 39.00 & 35.00 \\
		CUHK-2021 QuartzNet + CTC + Meta Learning + SAT \cite{wang2021improved} & 69.30 & 33.70 & 30.50 \\
		CUHK-2021 DNN + DCGAN + LHUC-SAT \cite{jin21_interspeech} & 61.42 & 27.37 & 25.89 \\
		CUHK-2021 DA + SBE Adapt + LHUC-SAT \cite{geng21b_interspeech} & 59.83 & 27.16 & 25.60 \\
		\textbf{TDNN + VAE-GAN + LHUC-SAT (Sys. 10, Tab. \ref{tab:overall})} & \textbf{57.31} & 28.53 & 27.78 \\
		\hline
		\hline
  	\end{tabular}
	}
\end{table}

\subsection{Performance after LHUC-SAT}
To model the speaker variability in the original and augmented data, LHUC-SAT speaker adaptation is performed (last 5 col. Tab. \ref{tab:overall}). 
Several trends can be observed: 
\textbf{1)} LHUC-SAT can bring up to 2.72\% (Sys. 4, last col.) absolute (8.89\% relative) WER reduction compared with those without adaptation (Sys. 4, col. 11);
\textbf{2)} After applying LHUC-SAT, the VAE-GAN based DA approach consistently outperforms speed perturbation by up to 1.27\% absolute (\eg Sys. 4 \textit{vs.} 2, last col.) (4.36\% relative) WER reduction;
\textbf{3)} Similarly the VAE-GAN based DA consistently outperforms the non-VAE GAN approach by up to 0.62\% absolute (\eg Sys. 4 \textit{vs.} 3, last col.) (2.17\% relative) WER reduction;
\textbf{4)} After LHUC-SAT speaker adaptation, the VAE-GAN augmented TDNN Sys. 10 gives the lowest average WER of 27.78\% among all systems in Tab. \ref{tab:overall}. 
This is further contrasted with recently published results on the same task in Tab. \ref{tab:comparison}.

\subsection{An Ablation Study}
is conducted to analyze the impact from the three sources of supervision used in structured VAE-GAN training in Sec. \ref{sec:structured_vae_gan}: speaker IDs, phone labels and pre-trained Wav2vec 2.0 features. 
The detailed analyses of Tab. \ref{tab:ablation_study} suggest for both TDNN and Conformer systems, data augmented using VAE-GANs incorporating all three sources of supervision consistently produced the best performance on the ``VL'' subset (\eg Sys. 10 \textit{vs.} 6-9 and Sys. 15 \textit{vs.} 11-14). 
Similar trends are found when combining TDNN and Conformer systems via two pass rescoring \cite{cui22_interspeech} (Sys. 20 \textit{vs.} 16-19). 

\section{Conclusion}

This paper proposed VAE-GAN based data augmentation approaches for disordered speech recognition. 
Experiments on the UASpeech dataset suggest improved coverage in the augmented data and model generalization is obtained over the baseline speed perturbation and non-VAE GAN based approaches, particularly on impaired speech of very low intelligibility.
Future research will improve the controllability of VAE-GAN models during data generation and application to non-parallel disordered speech data.  

\newpage
\bibliographystyle{IEEEtran.bst}
\bibliography{intro_bg.bib}

\end{document}